\documentclass[aps,prl,twocolumn,showpacs,preprintnumbers,amsmath,amssymb,floatfix]{revtex4}
\usepackage{graphicx}

\begin{document}

\title{Phase field approach for modeling intracellular dynamics}
\author{Julien Kockelkoren, Herbert Levine, and Wouter-Jan Rappel}

\address{Department of Physics and Center for Theoretical Biological
 Physics, University of California, San Diego, La Jolla, CA 92093-0319 
 }

\begin{abstract} 
We introduce a phase field approach for diffusion inside and outside
a closed cell with damping and with source terms at the interface.
The method is compared to
exact solutions (where possible) and the more traditional finite element
method. It is shown to be very accurate, easy to implement and
computationally inexpensive. 
We apply our method to a recently introduced model for
chemotaxis by Rappel et al. [Biophys. J. {\bf 83}, 1361 (2002)]. 
\end{abstract}           
 
\maketitle

%\section{Introduction}
In dealing with free boundary problems, the so called phase field
approach \cite{pf} appears as a method of choice. It has successfully been 
applied to various problems ranging from dendritic solidification
\cite{kr98}, viscous fingering \cite{pf-hs} and crack propagation \cite{pf-crack}.
 In the spirit of time-dependent
Ginzburg-Landau models, the
method avoids the tracking of the interface by introducing an auxiliary
field that locates the interface and whose dynamics is coupled to the
other physical fields through an appropriate set of partial
differential equations.
In comparison to the more traditional boundary
integral methods, the method is much simpler to implement numerically.

In this Brief Report we introduce a phase field model for
intracellular dynamics i.e.\ diffusion
inside and outside a  stationary,
closed domain with source terms at the interface.
We apply the method to a recently introduced model
for the response of a Dictyostelium amoeba \cite{bill} following stimulation with
the chemoattractant cAMP \cite{rltl02}. In \cite{rltl02}, due to the
need to use a finite element method the numerical implementation of the model
was limited to two
space dimensions and the cells were treated as disks. 
As we will see below, the phase field method is capable of faithfully 
capturing no-flux boundary conditions.
Thus, it becomes feasible to investigate more
realistic cell shapes in three dimensions. 

Before we introduce our approach, we like to point out some 
possible extensions of our methodology. 
The phase field approach can be easily modified to 
include problems where the domain boundary is not stationary.
For example, force generation on cell membranes, leading to 
shape changes, can be incorporated within the phase field approach
in a straightforward manner.
This would require adding an additional equation for the phase field,
but does not require the explicit calculation of a boundary.
We should stress that attempting to model this type of problem using 
conventional techniques, with explicit boundary tracking,
become quite cumbersome.
%Another advantage is that
%some dynamics of the cell interface can be easily incorporated, which
%is in fact one of the  Awaiting further results, we here
%focus on the case of static cell shapes. 

%\section{Diffusion inside the cell}
%For the sake of simplicity we first present the main features of our
%method in a one dimensional ``cell''. 
%An advantage here is that we can
%easily compare the results to finite-differnce en exact solution. 
Let us first introduce the most salient ingredients of our method.
Our purpose is to describe the situation where some
chemoattractant diffuses only inside the cell, i.e. its concentration
$c$ obeys the diffusion equation inside a  closed
domain:
$$
\frac{\partial c}{\partial t} = D \vec{\nabla}^2 c 
$$
and satisfies zero-flux boundary conditions: 
$$
\hat{n} \cdot \vec{\nabla}c = 0 
%\frac{\partial c}{\partial r}\Big|_{r=r_0}=0.
$$
As a phase field we take a function that takes the value $\phi_{in}=1$
inside the cell and $\phi_{out}=0$ outside the domain and varies smoothly
across the interface. A possible choice in one dimension  for a cell
between $x=-a$ and $x=a$ is:
\begin{equation}
\phi(x)=\frac12 + \frac 12 \tanh((a - |x|)/\xi)  
\label{phi-x}
\end{equation}
The variable $\xi$ denotes the interface width. 
We then define a field $u$ that is to obey the equation:
\begin{equation}
\phi \frac{\partial u}{\partial t} = \vec{\nabla} \cdot \left[
D \phi \vec{\nabla} u \right]
\label{pf-2d}
\end{equation}

Our claim is now that inside the domain 
the field $u$ behaves very similarly to $c$. 
Let us therefore first show, for simplicity in one dimension,
that in the thin interface limit one
recovers the no-flux boundary condition.   
Integration of (\ref{pf-2d}) over the interface yields:
$$
\int_{a-\xi}^{a+\xi}dx  \phi \frac{\partial u}{\partial t} 
 \approx - D \frac{\partial u}{\partial x}\Big|_{x=a-\xi}
$$
% - D \frac{\partial^2 u}{\partial \theta^2} \right]
since $\phi(a-\xi)\approx 1$ and $\phi(a+\xi)\approx 0$.
From this we deduce:
$$
\frac{\partial u}{\partial x}\Big|_{x=a} \sim \xi
$$
and thus, in the sharp interface limit $\xi \rightarrow 0$, the reflective
boundary conditions are recovered. 
%
%Moreover, rewriting Eq.\ (\ref{pf-1d}) as
%$$
%\frac{\partial u}{\partial t} =
%D \frac{\partial^2  u}{\partial x^2} +  D \frac{\phi'}{\phi} \frac{\partial
%  u}{\partial x }  
%$$
%with $\phi'=\frac{\partial \phi}{\partial x}$,
It is also clear that inside the domain where $\phi$ is constant $u$ satisfies the
diffusion equation.

%We also note that the factor $\phi$ on the left hand side makes the
%approximation at least two times better than without it. 

Another interesting property of Eq.\ (\ref{pf-2d}) is that the
 quantity $\int \phi u \,d\vec{x}$, which can be interpreted as the total
 concentration inside the cell, is conserved under the
 dynamics. Thus,
even if the field $u$ may become non-zero {\em outside} the cell (and
 indeed it does), the total
concentration {\em inside} the cell remains constant. In fact, the
asymptotic solution of Eq.\ (\ref{pf-2d}) (together with phase field
 Eq.\ (\ref{phi-x})) is $u(x)=A$ where the
 constant $A$ is determined through  $2 a A = \int dx \phi
 u(t=0)$.
The real asymptotic solution is of course $u(x) = \tilde{A}=\int_{-a}^{a} dx
   u(t=0) /(2 a)$.  
Because $\phi=1$ inside the domain and $\phi=0$ outside of it,
the error $\cal E$ in the asymptotic solution depends solely on  
the
 amount of concentration initially near the interfaces: $ {\cal E} \approx
 \int_{\pm a -\xi}^{\pm a +\xi} 
 dx  \phi u(t=0) / 2 a $.
Since $\phi$ is an antisymmetric function around the interface, 
the error is minimal if we take in the phase field approach an
 initial condition $u(t=0)$ which is locally symmetric around the boundary.

This raises the following important point. While the solution $u$
outside the cell is not of physical interest, it is essential to keep
track of it. In practice we solve (\ref{pf-2d}) where  the phase field
$\phi$ exceeds 
a small threshold $\delta$ (typically $\delta \sim 10^{-8}$).

As a first test for our model we now solve Eq.~(\ref{pf-2d}) numerically 
in two space dimensions.
For the phase field we take here: 
\begin{equation}
\phi(r)=\frac12 + \frac 12 \tanh((r_0-r)/\xi)  
\label{phi-rho}
\end{equation}
where $r$ is the radius in polar coordinates.
In the case of a radially symmetric initial condition we can compare the
field $u$ to the analytic solution $v$ which is expressed in terms of
the Bessel functions
$$
v(r,t) = \sum_n a_n J_n(\lambda_n r) \textrm{e}^{-\lambda_n^2 t}
$$
where $\lambda_n$ is defined as the smallest number for which
$J_n(\lambda_n a)=0$ and $a_n$ is the projection of the initial
condition on the set of Bessel functions.  
We also compare both fields $u$ and $v$ with the solution $w$ of
a finite element method obtained with MATLAB's PDE Toolbox (The
Mathworks, Natick, MA). 

\begin{figure}[h!]
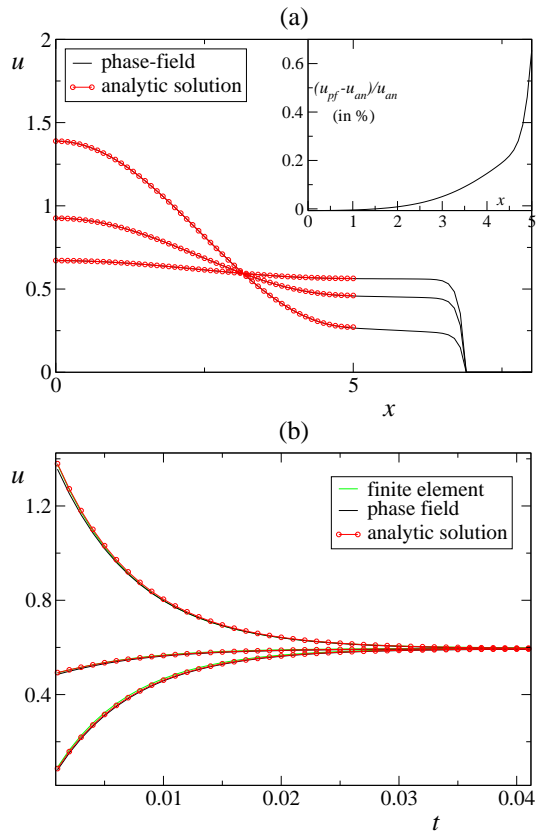

\includegraphics[width=7.cm,clip]{fig1a.eps}
\includegraphics[width=7.cm,clip]{fig1b.eps}
\caption{\protect\small Comparison of our phase field model with both
  analytic solution and finite element method.  
The equations are solved on a  grid of size $201 \times 201$ with
Euler's method, with $D=250$, $r_0=5$, $\Delta x=0.1$, and $\Delta t=5\cdot 10^{-6}$. 
The initial condition is $c=1+\cos(\pi  r/r_0)$.
 (a) Spatial profile at
 respectively at $t=0.004$, $t=0.01$ and $t=0.02$  of the phase field and
  analytic solution. Since the curves are hardly distinguishable, we
  mark the analytic solution by small circles and plot the difference
  of the profiles at $t=0.01$ in the inset. The relative error of the phase
  field is seen to be smaller than 1\%, being maximal at the boundary. 
(b) Time evolution of
  the concentration at $r=5$, $r=3.37$ and $r=1.66$. The analytic
  solution is again marked by small circles.}
\label{fig1}
\end{figure}

In Fig.~\ref{fig1}a we show the spatial profile of the phase-field
together with the exact solution at a given time.
It can be seen that the agreement is excellent, with a relative error
of less than $1 \%$ (see insert).
In Fig.~\ref{fig1}b we show the fields $u$,$v$ and $w$ at three different
points. Again the agreement is excellent. 

In view of these promising results, let us now consider diffusion in
presence of a production term and damping.
%\section{Including a force term}
Including a source term at the interface is relatively easy. We add to
the right hand side of (\ref{pf-2d}) a term $b (\vec{\nabla} \phi)^2 / \int d\vec{x}\,
(\vec{\nabla} \phi)^2$. The factor
$(\vec{\nabla} \phi)^2$  ensures that it only acts at the
interface and the denominator is a normalization factor.
 To compensate for the production we also include a damping
term of the form $-\mu \phi u$. The equation of motion then becomes:

\begin{equation}
\phi \frac{\partial u}{\partial t} = \vec{\nabla} \cdot \left[
D \phi \vec{\nabla} u \right] - \mu \phi u + b \frac{(\vec{\nabla}
  \phi)^2}{\int d\vec{x}\, (\vec{\nabla} \phi)^2}
\label{pf-2d-prod} 
\end{equation}

\begin{figure}[h!]
\includegraphics[width=7.cm,clip]{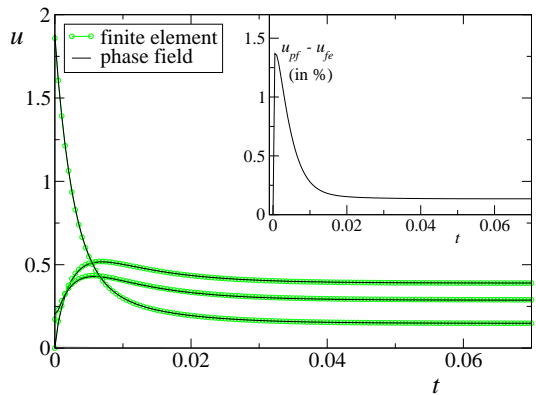}

\caption{\protect\small Comparison of our phase field model with 
  finite element method with source and damping. We have taken the same system size, initial conditions and parameter values as in Fig.\ \ref{fig1}, and now 
  $b=1000$ and $\mu=50$. 
 We show here the time evolution of
  the concentration at $r=5$, $r=4.05$ and $r=0.86$. The curves are
  again hardly distinguishable. We plot the error $u_{pf}-u_{fe}$ for
  $r=5$ (the worst point of Fig.1a) as a function of time. 
As one can see in the insert, the
  error goes initially rapidly up to around 1.5~$\%$, but decreases then
  to around $6
  \cdot 10^{-2}\%$. }
\label{fig2}
\end{figure}

We have compared the phase field method with these two supplementary
ingredients with the finite element method, again in the case of a
two-dimensional circular cell with $r_0=5$. As can be seen in
Fig.~\ref{fig2} the result is excellent. 

We now use our phase field method to solve a biological model
for the response of a Dictyostelium amoeba following stimulation with
the chemoattractant cAMP \cite{rltl02}. In recent experiments the
establishment of an asymmetry within a few seconds after a rise of
extracellular cAMP was demonstrated. The cAMP however diffuses rapidly
around the cell and the applied signal is several orders of magnitude
larger than the value required to elicit a response. This strongly
suggests the presence of  
an inhibitory mechanism that suppresses responses at the back.

In \cite{rltl02} an abstract model for the initial response of the cell
to the chemoattractant was
proposed: it was supposed that the membrane can be characterized 
in terms of three states: quiescent (with density $\rho_q$), activated
(with density $\rho_a$) and
inhibited (with density $\rho_i$). Initially the entire membrane is in the 
quiescent state.
As the cAMP
reaches the front of the cell the membrane becomes activated at rate
$\alpha [{\rm cAMP}]$ and an
inhibitor, in \cite{rltl02} suggested to be cGMP, is produced (at rate
$\sigma_g \rho_a$). The
inhibitor diffuses toward the back of cell where it turns the
membrane from quiescent to 
inhibited with rate $\beta_r [{\rm cGMP}]$. Furthermore both activated
and inhibited state decay spontaneously to the quiescent state at
rates $\delta$ and $\beta_f$ respectively.  
The equations for the membrane state variables are thus:
\begin{eqnarray}
\frac{\partial \rho_q}{\partial t} & = & - \alpha c \rho_q + \beta_f
\rho_i - \beta_r g \rho_q \\
\frac{\partial \rho_a}{\partial t} & = & \alpha c \rho_q -\delta \rho_a \\
\frac{\partial \rho_i}{\partial t} & = & - \beta_f
\rho_i + \beta_r g \rho_q + \delta \rho_a
\end{eqnarray}
 
The reactants cGMP and cAMP diffuse respectively inside and outside
the cell. At the membrane they satisfy zero-flux boundary
conditions. There is an source term for the cGMP field that
accounts for the production of cGMP at the interface. Both cAMP and
cGMP fields are damped at rates $\mu_c$ and $\mu_g$. 
The phase-field method is thus well suited to solve the dynamical
equations for the cAMP and cGMP concentrations. For the phase field
corresponding to the cAMP (which diffuses outside the cell) we simply
take the complement of $\phi$ given by Eq.\ (\ref{phi-rho}): $\phi_c = 1 -
\phi$.  The equations for the membrane variables are solved on all
lattice sites where $(\vec{\nabla} \phi)^2$ exceeds a certain threshold,
namely $10^{-4}$.  

\begin{figure}[h!]
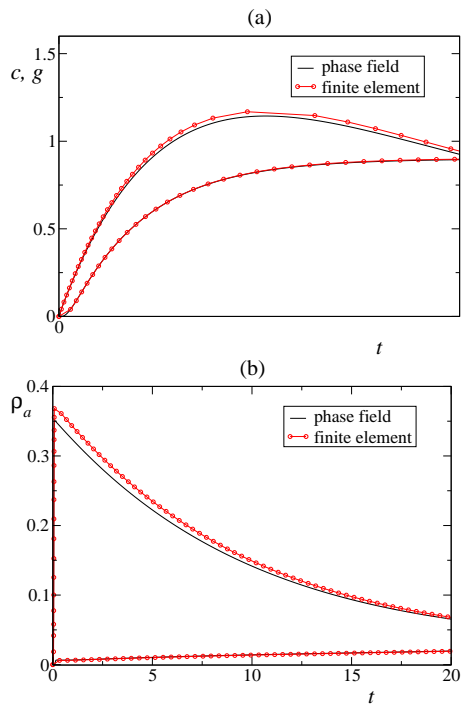

\includegraphics[width=6.cm,clip]{fig3ab.eps}
\includegraphics[width=6.cm,clip]{fig3c.eps}
\caption{Comparison of our phase field model with 
 finite element method in the chemotaxis model.  
 (a) cGMP concentration at front (upper curves) and cAMP at back
 (lower curves) of the cell as a function
 of time
  (b) state variables $\rho_a$ at front and back of the cell as a
 function of time.
In the finite element method the grid consists of 216 nodes inside the cell
 (of which 40 on the interface) and 543 on the outside (of which again
 40 on the interface). The grid and also the mass and stiffness
 matrices were generated by MATLAB's PDE Toolbox (The Mathworks, Natick, MA) 
after which the
 equations of motion are integrated with a Fortran code. The time step
 is taken to be $\Delta t = 2 \cdot 10^{-6}$.   
In the phase field method the equations of motion are integrated on a
 $151 \times 151$ grid, the lattice spacing thus being $\Delta x=0.2$. 
Here the time step is taken to be $\Delta t = 10^{-5}$.  We have taken
 the following parameter
 values: $\alpha= 4$ ,  $\beta_f= 0.01$,   
$\beta_r = 0.533$, $\delta = 0.1$,
$\mu_c = 0$,
$\mu_g = 0.12 $, and
$\sigma_g = 60000 $.
}

\label{fig3}
\end{figure}

We now present a comparison of the phase field approach
and the results obtained with a finite element method in
\cite{rltl02}. A circular cell of diameter 10 $\mu$m is placed in a
square domain of 30 $\mu$m by 30 $\mu$m. The diffusion constant of
cAMP and cGMP was taken to be identical: $D_c=D_g=250 \mu$m$^2$/s.
The values of the other parameters can be read in the caption of 
Fig.\ref{fig3}.

In order to mimic the asymmetric cAMP stimulus we maintain the cAMP
concentration at a value well above threshold at the upper left corner
of our domain. 
As expected from our earlier results the agreement of the cAMP fields,
that solely diffuse around the cell, is excellent, see
Fig.~\ref{fig3}a. 
We observe a slight discrepancy for the cGMP field (also
Fig.~\ref{fig3}a), 
which grows with
time. This is related to the larger difference 
(of around 10\%) for the membrane variables, the production of cGMP
being proportional to $\rho_a$.
The origin of the discrepancy might
lie in the way the state variables are calculated: on a ring of finite
width in the phase field model, and on 40 points on the interface in
the finite element method. 
At any rate, since the model is of a quite abstract nature and since its
predictions are only qualitative, a detailed investigation is beyond
the scope of this paper. 

From a computational perspective, we note that in the phase field
 method the CPU time required is  linear in the number of lattice
 sites whereas it is quadratic in the number of nodes in the 
 finite element method. One thus expects the phase field method to be
 much faster.  However this naive result must be taken with some caution as in the
 finite element method the nodes are not distributed uniformly in
 space. To resolve some particular region in space with high accuracy
 we are thus obliged to take a small lattice spacing in the phase
 field method, such that the number of lattice points exceeds  the
 number of nodes. Other parameters that will affect the accuracy of
 both methods are the time step $\Delta t$ and the integration
 algorithm. Again a detailed comparison  is beyond the scope of the Report.
In practice it turns out that the phase field method where the lattice
 spacing is equal to the minimal distance between two nodes of our
 finite element method is about one order of magnitude faster.

\begin{figure}[htb]
\vspace{0.5cm}
\includegraphics[width=7.cm,clip]{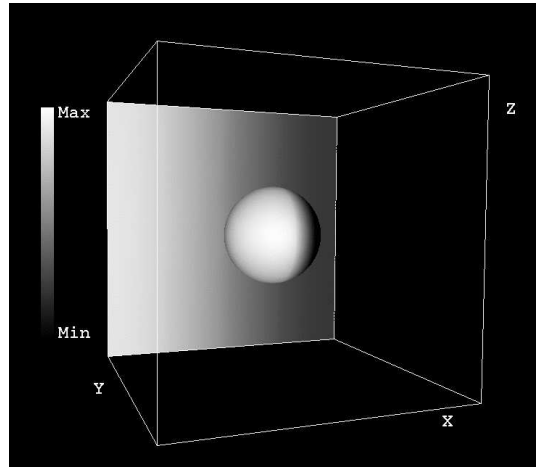}
\caption{The cAMP concentration,
  represented by gray levels (where white corresponds to large
  concentrations and black to low ones) is for visual simplicity
shown at the back of the box and is similar to the cAMP field
surrounding the cell.
The activity density of the membrane is shown on the sphere,
  also represented by gray levels. It is seen that whereas the cAMP
  has reached the back of the cell, it has not become activated.     
The equations of motion are integrated on a
 $61 \times 61 \times 61$ grid, the lattice spacing thus being $\Delta x=0.5$. 
Here the time step is taken to be $\Delta t = 10^{-4}$. These results
have been obtained for the same parameters as in Fig.~\ref{fig3}. Only the
  production term has been increased in order to account for the
  increase in membrane surface. 
 }
\label{fig4}
\end{figure}

Finally, we turn to the chemotaxis model in three space dimensions.  For the
sake of simplicity we take a spherical cell, i.e. with a phase field
like in Eq.\ (\ref{phi-rho}) but where $r$ is now the radius in spherical
coordinates. We consider again a cell of radius $r_0=5 \mu m$, in a
box of dimensions $30 \times 30 \times 30 \mu m$. Now the stimulus is
applied by setting the cAMP initially well above threshold at one side
of the box, here taken to be $x=-15 \mu m$. 
As is illustrated in Fig.~\ref{fig4}, the phenomenology of the two
dimensional case is reproduced here. As the cAMP front progresses
toward the back of the cell, only the front of the cell is
activated. 

In conclusion we have proposed a phase field model for intracellular
dynamics.
 Our method is shown to be
very accurate, easy to implement and computationally inexpensive. 
Another advantage lies in its much greater flexibility with respect to
other methods, like the finite element method. 
We revisited a chemotaxis model and, when considering
three dimensional cells, we reach the same conclusions as in \cite{rltl02}
for two-dimensional cells. 
Whereas we have
restricted ourselves to circular and spherical domains, the extension
to other geometrical forms poses no major problems, the only task
being to generate a phase field $\phi$. 
Even better, our approach can easily be extended to deal with 
non-stationary boundaries. This situation arises in a multitude of 
biological problems. For example, the Dictyostelium cells change their
shape continuously during chemotaxis. Another example are shape
transformations observed in vesicles \cite{seifert}. 
In these cases, the phase field
becomes a dynamic
variable, that evolves under the appropriate Ginzburg-Landau type of
equation.  
We are currently working along these lines of research.  

%On a final note, we believe that our method, particularly in light 
%of its relative ease of implementation, can also be of great interest
%outside the field of biological modeling. We thus hope that the present
%work will trigger similar studies of diffusion problems in complex 
%geometries, possibly with moving boundaries,
%in other fields. 

This work was supported by the NSF sponsored Center for Theoretical 
Biological Physics (grants  PHY-0216576 and 0225630) and the NSF
Biocomplexity program (grant MCB 0083704).

\end{document}